\documentclass{article}

\usepackage[final, nonatbib]{neurips_2021}

\usepackage[utf8]{inputenc} %
\usepackage[T1]{fontenc}    %
\usepackage{hyperref}       %
\usepackage{url}            %
\usepackage{booktabs}       %
\usepackage{amsfonts}       %
\usepackage{nicefrac}       %
\usepackage{microtype}      %
\usepackage{xcolor}         %

\usepackage{graphicx}
\usepackage{multirow}
\usepackage{booktabs}
\usepackage{floatrow}%
\usepackage{blindtext}
\usepackage[export]{adjustbox}
\usepackage[ruled,vlined]{algorithm2e}
\usepackage{amsmath}
\usepackage{epsfig}

\usepackage{caption}
\usepackage{subcaption}
\usepackage{siunitx}
\newfloatcommand{capttabbox}{table}[\captop][\FBwidth ]

\title{Tracking Urbanization in Developing Regions with Remote Sensing Spatial-Temporal Super-Resolution}

\author{%
    Yutong He\thanks{Equal contribution.} \qquad William Zhang\footnotemark[1] \qquad Chenlin Meng \\ \textbf{Marshall Burke \qquad David B. Lobell \qquad Stefano Ermon}\\
    Stanford University \\
  \{\tt kellyyhe, wxyz, chenlin, ermon\}@cs.stanford.edu \\
  \{\tt mburke, dlobell\}@stanford.edu \\
}

\begin{document}
\maketitle

\begin{abstract}
\label{abstract}

Automated tracking of urban development in areas where construction information is not available became possible with recent advancements in machine learning and remote sensing. Unfortunately, these solutions perform best on high-resolution imagery, which is expensive to acquire and infrequently available, making it difficult to scale over long time spans and across large geographies. In this work, we propose a pipeline that leverages a single high-resolution image and a time series of publicly available low-resolution images to generate accurate high-resolution time series for object tracking in urban construction. Our method achieves significant improvement in comparison to baselines using single image super-resolution, and can assist in extending the accessibility and scalability of building construction tracking across the developing world.

\end{abstract}

\section{Introduction}
Accurate measures of building construction and urban development are an important factor for measuring and understanding economic development and population growth.  Such measures have thus become important in informing a range of government policy decisions, including how and where to target public service delivery.  However, many regions may lack the resources to systematically measure such development over large geographies or over time \cite{burke2021using}.
In the meantime, satellite imagery has been proven to be useful in many human development applications such as poverty prediction, infrastructure measurement, and tracking sources of pollution, especially in developing areas where survey data or labeled data is difficult to obtain \cite{ayush2020generating, burke2021using, lee2021scalable}. Compared to traditional survey-based methods, remote sensing approaches can in principle repeatedly observe large areas at potentially low cost, offering the ability to scale measurements of key development outcomes. %

Recently, SpaceNet released a satellite image dataset including 4m ground sample distance (GSD) imagery and labels for tracking building construction in rapidly urbanizing areas, which attracted many solutions from the deep learning community \cite{van2021spacenet}.
These algorithms require high-resolution (HR) imagery to achieve their best performance.
However, HR imagery is expensive and 
captured infrequently, limiting the scalability of those methods \cite{burke2021using}. On the other hand, lower-resolution remote sensors provide publicly available imagery and shorter revisit periods \cite{jerven2017much}. Unfortunately, many objects of interests (eg. residential houses) will be too small to be visible in lower-resolution imagery.

Spatial-temporal super-resolution of satellite imagery has shown great potential for generating realistic and accurate HR images \cite{he2021spatialtemporal}. Unlike single image super-resolution, it leverages an auxiliary HR reference image and introduces additional temporal information to improve the image generation quality.

Inspired by this success, we propose a pipeline that uses a spatial-temporal super-resolution model which uses one HR image and a times series of low-resolution (LR) images to generate the target HR time series, which is then input into an object tracker to monitor construction in the area of interest (AOI).
To examine the effectiveness of our algorithm, we collect a LR dataset corresponding to SpaceNet 7 with Landsat 8 (30m GSD) and conduct quantitative and qualitative analysis on this paired dataset.
We show that spatial-temporal super-resolution greatly improves the performance of the object tracker compared to single image super-resolution. We anticipate this method can help expand the accessibility and temporal scalability of object tracking for assessing urbanization in developing regions.
\section{Related Work}
Object tracking in satellite image time series has been shown to work with high-resolution imagery \cite{meng2012object}. Challenges and datasets such as \cite{van2021spacenet, gupta2019xbd, christie2018functional} include high-resolution time series of satellite imagery, temporal information, and labels with changes over time, with the goal of tracking objects over time. However, the satellite imagery in these datasets come from paid sources, making it expensive to conduct large scale experiments requiring considerable amounts of imagery and is therefore less accessible to regions lacking resources.

Deep models have achieved state-of-the-art performance in single image super-resolution \cite{srcnn, srgan, dbpn}.
However, previous work has shown that these models are less applicable for satellite imagery at lower resolution and larger scale factors, especially in the context of object detection tasks \cite{effectofsronsatellite} .
In recent years, fusion models that blend signals from two remote sensing devices have shown great potential in improving spatial detail in generated images \cite{starfm, estarfm, unmixing, cganfusion}. He et al. \cite{he2021spatialtemporal} proposed a spatial-temporal super-resolution model that can produce high quality accurate HR images. We follow \cite{he2021spatialtemporal} with additional training objectives that improves the performance of the object tracking task.

\section{Method}
\label{method}

\begin{figure}
    \centering
    \includegraphics[width = 0.8\textwidth]{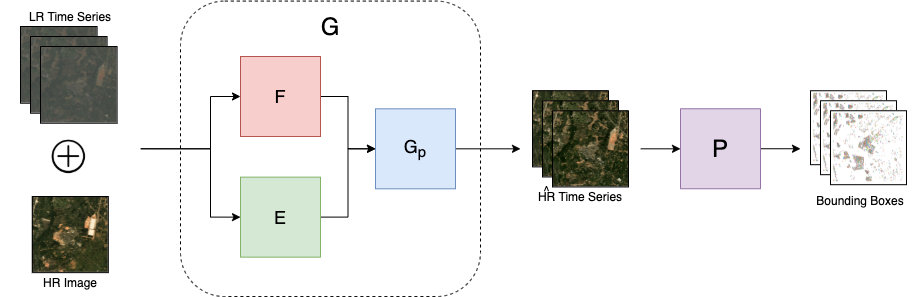}
    \caption{Our proposed pipeline uses a time series of low-resolution images and a single high-resolution image to generate a time series of high-resolution images. These are then input into an object tracker to generate building labels for each of the images in the time series.}
    \label{fig:pipeline}
\end{figure}

The aim of this work is to track building constructions in areas with rapid urbanization using a time series of low-resolution (LR) satellite images and a single high-resolution (HR) satellite image. With an object tracker, we leverage the abundance of LR imagery to enhance geographical and temporal scalability and the high precision details provided by the HR imagery to improve the model accuracy of tracking building constructions.

Let $I_{lr}^{(t)} \in \mathbf{R}^{C\times H_{lr}\times W_{lr}}$ and $I_{hr}^{(t)} \in \mathbf{R}^{C\times H\times W}$ be a LR and a HR satellite image of the area of interest (AOI) at time $t$. $C, H_{lr}, W_{lr}, H, W$ represent the number of bands, the height and width of the LR images and the height and width of the HR images. At training time, we have access to paired satellite image time series $\{I_{lr}^{(t)}\}_{t=t_0}^{T}$ and $\{I_{hr}^{(t)}\}_{t=t_0}^{T}$. At inference time, only the LR time series $\{I_{lr}^{(t)}\}_{t=t_0'}^{T'}$ and the most recent HR image $\{I_{hr}^{(T')}\}$ are provided.

To achieve this goal, we first use a spatial-temporal super-resolution model $G$ to obtain the estimated HR image time series $\{\hat{I}_{hr}^{(t)}\}_{t=t_0'}^{T'}$, and then apply a object tracker $P$ to produce the predicted building footprints. At each step, we denote the LR and HR image at target time $t$ as $I_{lr}^{(t)}$ and $I_{hr}^{(t)}$ and another HR image at time $t'\neq t$ as $I_{hr}^{(t')}$.

\paragraph{Spatial-Temporal Super-Resolution}
We follow \cite{he2021spatialtemporal} to perform the spatial-temporal super-resolution task. \cite{he2021spatialtemporal} is a conditional pixel synthesis model which consists of an image feature mapper $F$, a positional encoder $E$ and a pixel synthesizer $G_p$.

The image feature mapper $F$ first resizes $I_{lr}^{(t)}$ to $H\times W$ and takes the band-wise concatenation $I_{cat}^{(t)} = \text{concat}(I_{lr}^{(t)}, I_{hr}^{(t')}) \in \mathbf{R}^{2C\times H\times W}$ as its input. Then $F$ extracts the image features from $I_{cat}^{(t)}$ and map the features to each pixel in the $H\times W$ coordinate grid $X$ using convolutional layers, self-attention modules and transpose convolutional layers.
The positional encoder $E$ calculates the Fourier features of coordinate $(x,y,t)$ and the spatial coordinate embedding of $(x,y)$ for all $(x,y)$ in $X$. We denote the resulting positional encoding of the entire image as $E(X,t)$.
The pixel synthesizer $G_p$ uses the image feature $F(I_{cat}^{(t)})$ extracted from $F$ and the positional encoding $E(X,t)$ computed from $E$ to predict the pixel value at each $(x,y)$ in the coordinate grid $X$. The final estimated HR image can be calculated as
$\hat{I}_{hr}^{(t)} = G(X,t|I_{lr}^{(t)}, I_{hr}^{(t')}) = G_p(F(I_{cat}^{(t)}), E(X,t))$.

Following \cite{he2021spatialtemporal}, we use L1 loss to encourage the outputs to respect the input image structures, and train a discriminator $D$ to include conditional GAN loss in the optimization process to create realistic outputs. In addition, we also introduce a deep perceptual similarity criterion, LPIPS \cite{zhang2018perceptual} to further improve the generation quality. With hyperparameters $\lambda_1, \lambda_2$, the objective function is

\begin{align*}
G^* &= \arg\min_G\max_D \mathcal{L}_{cGAN}(G,D) + \lambda_1 \mathcal{L}_{L_1}(G) + \lambda_2 \mathcal{L}_{LPIPS}(G)\\
\mathcal{L}_{L_1}(G) &= \mathbb{E}[||I_{hr}^{(t)} - \hat{I}_{hr}^{(t)}||_1]\\
\mathcal{L}_{LPIPS}(G) &= \mathbb{E}[\text{LPIPS}(I_{hr}^{(t)}, \hat{I}_{hr}^{(t)})]\\
\mathcal{L}_{cGAN}(G,D) &= \mathbb{E}[\log D(I_{hr}^{(t)}, X, I_{lr}^{(t)}, I_{hr}^{(t')})] + \mathbb{E}[1-\log D(\hat{I}_{hr}^{(t)}, X, I_{lr}^{(t)}, I_{hr}^{(t')})]
\end{align*}

\paragraph{Object Tracker}
We choose the winning model \cite{spacenetwinning} of the SpaceNet 7 Challenge \cite{van2021spacenet} as the object tracker $P$. Given the generated HR time series $\{\hat{I}_{hr}^{(t)}\}_{t=t_0'}^{T'}$ from $G$, we use HRNet \cite{hrnet}, an image semantic segmentation model to determine a rough estimation of the locations of the building constructions in each image of the time series. Then we perform a spatial-temporal collapse post-processing to the predicted polygons: temporal collapse first compresses all areas that have changed in at least one time step into a single probability map; spatial collapse then determines the changing time step for each polygon in the probability map. The combination of the temporal and spatial collapse produces the final prediction result for the AOI.
\section{Experiments}
\label{experiment}

\subsection{Datasets}
We use the Multi-Temporal Urban Development SpaceNet (MUDS, also known as SpaceNet 7) dataset \cite{van2021multi} as our HR data source, which consists of 4m GSD satellite imagery collected from Planet’s global monthly
basemaps. The released dataset contains images and building footprint labels between 2017 and 2020 in 60 locations across the globe, which were selected to be geographically diverse and display dramatic changes in urbanization. Each location has approximately 24 images (one per month) with dimensions of $1024 \times 1024$ pixels, corresponding to an area of $\approx 18$ \si{km^2}.

We collect corresponding low-resolution RGB imagery from Landsat 8 (30m GSD) using Google Earth Engine \cite{gorelick2017google}. Images were acquired in the same month and location as the corresponding image in the SpaceNet 7 dataset. Similar to the SpaceNet 7 dataset \cite{van2021multi}, images containing an excessive amount of clouds or haze were fully excluded from the low-resolution dataset, thus reducing the number of available images. After collecting the low-resolution images, we pair them with the corresponding high resolution images. We randomly select 50 AOIs for the training set and 10 AOIs for the testing set. There are 635 pairs in the training set and 119 pairs in the testing set. For training, we pair each LR image with an HR image from another timestamp to create the input tuples; for test, we couple each LR image with the most recent HR image of the same location to generate the estimation HR image of the target time.

\subsection{Implementation Details}
\paragraph{Model Details} We follow the EAD configuration in \cite{he2021spatialtemporal} for the spatial-temporal super-resolution model. We choose $C = 3$, $H = W = 256$ and $\lambda_1 = 100, \lambda_2 = 10$. We train our model using Adam optimizer with learning rate $2\times 10^{-3}, \beta_0 = 0, \beta_1 = 0.99, \epsilon = 10^{-8}$ on NVIDIA Titan XP GPUs. The training takes 120 hours for the model to converge. At inference time, we use the generating by patch technique in \cite{he2021spatialtemporal} to generate the large $1024\times 1024$ images.

We also experiment with Pix2Pix \cite{pix2pix} architecture for $G$. The image encoder and decoder in Pix2Pix corresponds to $F$ and $G_p$ respectively and $E$ is omitted in this setting. Notice that the image encoder also takes the concatenated images %
as the input. The model is trained with $256\times 256$ images and inference with the same generating by patch technique to create the $1024\times 1024$ images.

The object tracker uses ImageNet pretrained weights. We then finetune it with 3X enlarged $512\times 512$ images using SGD optimizer with learning rate $0.01$. At inference time, we preprocess the input image into non-overlapping 3X enlarged $512\times 512$ patches and input them to the network.

\paragraph{Baselines}
Given the same object tracker, we compare our spatial-temporal super-resolution model with leading single image super-resolution solutions. Single image super-resolution methods generate the estimated HR images at the target timestamps given the corresponding LR images. We choose SRGAN \cite{srgan}, which is a widely used GAN based super-resolution model for satellite imagery, and DBPN \cite{dbpn}, which is a state-of-the-art super-resolution model for satellite imagery \cite{hybrid} as the baseline methods.

We also compare the generated images with the ground truth HR and LR images. The performance on the HR images serves as the upper bound of the model performance. We resize the LR images to match the input dimension of the model and the labeled bounding boxes for building constructions.

\paragraph{Evaluation Metrics} We evaluate our model with two types of metrics: one for assessing bounding box quality, another one for measuring object tracking ability. We report pixel accuracy (Acc), Intersection Over Union (IoU), and frequency-weighted IoU (FWIoU), which assess the predicted bounding box quality. We also evaluate our methods with the Tracking Score (TS) proposed by \cite{van2021spacenet}, which is a metric designed for measuring how well the proposal tracks the same buildings from month to month in satellite images.

\subsection{Results}

\begin{figure}[]
    \centering
    \includegraphics[width = 0.9\textwidth]{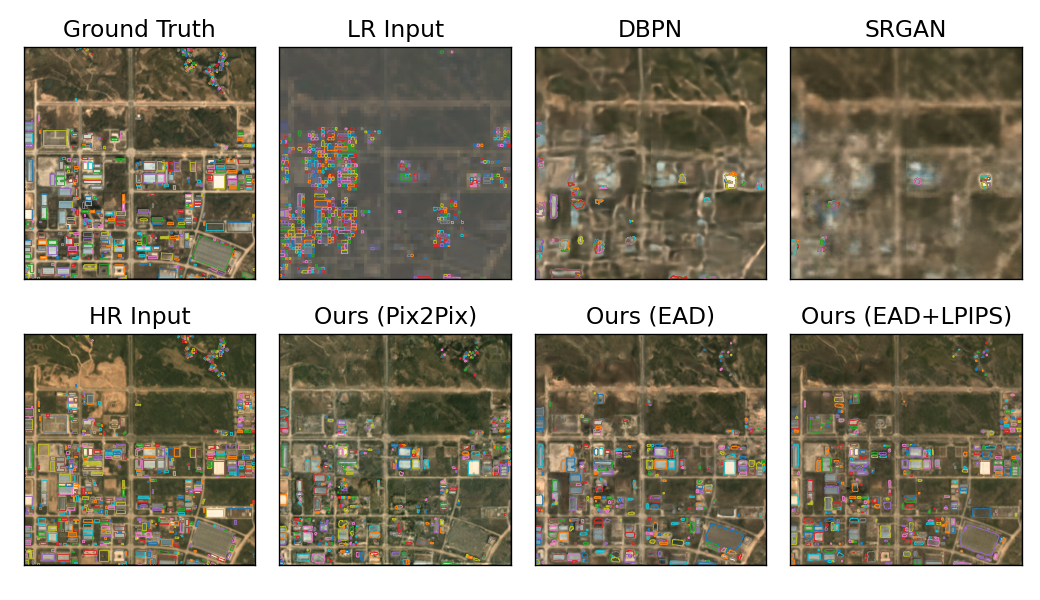}
    \caption{Object tracking results on different sources of imagery. Each colored box corresponds to an individual object. In comparison to single image super-resolution methods, our approach yields building construction tracking results that are more consistent with the ground truth polygons.}
    \label{fig:imagegrid}
\end{figure}

We evaluate the object tracker with two settings: (1) finetune the ImageNet pretrained object tracker on the original SpaceNet 7 images; (2) generate the training set separately using each method and finetune the ImageNet pretrained object tracker on the generated images. The first setting examines the generated images when a object tracker pretrained on HR dataset is provided, while the second setting allows us to explore a more practical scenario where the model pretrained on HR satellite images is not given. We use the same ground truth building footprints for both settings.

Figure \ref{fig:imagegrid} shows the qualitative results of the building construction tracking task on difference sources of images. Ground truth and LR input are the HR and LR image at the target time respectively, and HR input represents the most recent HR image that is input to the spatial-temporal super-resolution model. The object tracker performs significantly better on images generated with our method than on LR input or images generated by single image super-resolution models. Our pipeline is able to detect small building constructions that are only visible in HR images, and generate bounding boxes that are consistent with the ones produced on ground truth imagery. As shown in the figure, our method is also able to generate tracking results that is faithful to the LR input with accurate details learned from the HR input. Note that our model is nevertheless restricted by the same limitation as ~\cite{he2021spatialtemporal} and therefore when $t'<t$, the generation task becomes more challenging and we expect a similar performance degradation as a result.

Table \ref{tab:results} presents the quantitative results of the building construction tracking task. With spatial-temporal super-resolution, the object tracker achieves significant improvements in all metrics reported compared to baselines. Notice that when evaluated with TS, LR images and images generated by single image super-resolution are barely usable, while images generated by spatial-temporal super-resolution obtain comparable performance to ground truth HR images. Experiments in both settings agree with this conclusion. We also observe that EAD with LPIPS loss presents advantages in the majority of experiments, which shows the effectiveness of our proposed training objective.

\begin{table}[]
\centering
\begin{tabular}{c|cccc|cccc}
\toprule
\multirow{2}{*}{Test Image} & \multicolumn{4}{c|}{ImageNet + SpaceNet 7}             & \multicolumn{4}{c}{ImageNet + Target Domain}         \\ \cline{2-9}
                            & Acc$\uparrow$    & IoU$\uparrow$    & FWIoU$\uparrow$  & TS$\uparrow$ & Acc$\uparrow$ & IoU$\uparrow$    & FWIoU$\uparrow$  & TS$\uparrow$ \\ \hline
HR                          & 0.705 & 0.651 & 0.912 &  0.547                         & 0.705                    & 0.651 & 0.912 &  0.547    \\ \hline
LR                          & 0.534 & 0.471 & 0.823 &  0.043                         & 0.517                    & 0.486 & 0.879 &  0.066    \\
DBPN                        & 0.507 & 0.476 & 0.879 &  0.042                         & 0.515                    & 0.484 & 0.878 &  0.054    \\
SRGAN                       & 0.505 & 0.473 & 0.878 &  0.042                         & 0.514                    & 0.482 & 0.878 & 0.048  \\ \hline
Ours (Pix2Pix)                     & 0.621 & 0.581 & 0.897 & \textbf{0.324}                          & 0.616                    & 0.577 & 0.896 &\textbf{ 0.307}     \\
Ours (EAD)                         & 0.601 & 0.565 & 0.895 & 0.241                          & \textbf{0.628}                    & 0.583 & 0.896 & 0.288     \\
Ours (EAD+LPIPS)                   & \textbf{0.628} & \textbf{0.586} & \textbf{0.898} & 0.279                          & 0.626                    & \textbf{0.585} & \textbf{0.897} &  0.295    \\ \bottomrule
\end{tabular}
\caption{Building tracking performance using model trained with different settings. The results are grouped by test image domains. The first group is the ground truth HR imagery, the second group is the ground truth LR imagery and images generated by single image super-resolution, and the third group are images generated by spatial-temporal super-resolution using our proposed pipeline.}
\label{tab:results}
\end{table}

\section{Conclusion}

We propose a pipeline that uses a spatial-temporal super-resolution model to leverage the precise details in HR images and the cost effectiveness and availability of LR images for tracking urban development. This method significantly improves object tracker performance in new areas of interest compared to directly using LR images or single image super-resolution for image generation.

We expect that this framework will extend the ability of well-performing urbanization tracking models in developing regions lacking HR imagery and labeled data. Our method is more economical than methods using purely HR imagery, which allows for organizations with limited resources to perform studies over greater time spans and larger geographies. We hope our method can aid the scalability of tools for assessing building construction in the developing world and assist in decision making for sustainable urban development.

\section*{Acknowledgement}
This work was mainly funded by the IARPA SMART 2020-0072 project. This research is based upon work supported in part by the Office of the Director of National Intelligence (ODNI), Intelligence Advanced Research Projects Activity (IARPA), via 2021-2011000004. The views and conclusions contained herein are those of the authors and should not be interpreted as necessarily representing the official policies, either expressed or implied, of ODNI, IARPA, or the U.S. Government. The U.S. Government is authorized to reproduce and distribute reprints for governmental purposes not-withstanding any copyright annotation therein.

\bibliographystyle{plain}
\bibliography{reference}
\newpage

\end{document}